\documentclass[journal=jacsat,manuscript=article]{achemso}

\usepackage[version=3]{mhchem} 
\usepackage{xcolor}
\usepackage{soul}

\author{Petr Zhilyaev}
\affiliation{MIREA — Russian Technological University, Vernadsky Avenue 78, 119454 Moscow , Russia.}
\email{peterzhilyaev@gmail.com}

\author{Kirill Brekhov}
\affiliation{MIREA — Russian Technological University, Vernadsky Avenue 78, 119454 Moscow , Russia.}

\author{Elena Mishina}
\affiliation{MIREA — Russian Technological University, Vernadsky Avenue 78, 119454 Moscow , Russia.}

\author{Christian Tantardini}
\affiliation{Hylleraas center, UiT The Arctic University of Norway, PO Box 6050 Langnes, N-9037 Troms\o, Norway}
\alsoaffiliation{Department of Materials Science and NanoEngineering, Rice University, Houston, Texas 77005, United States of America}
\email{christiantantardini@ymail.com}

\title{Ultrafast Polarization Switching in BaTiO$_3$ Nanomaterial: \\ Combined DFT and Coupled Oscillator Study}

\abbreviations{IR,NMR,UV}

\begin{document}


\begin{tocentry}
\begin{center}
\includegraphics[width=1\textwidth]{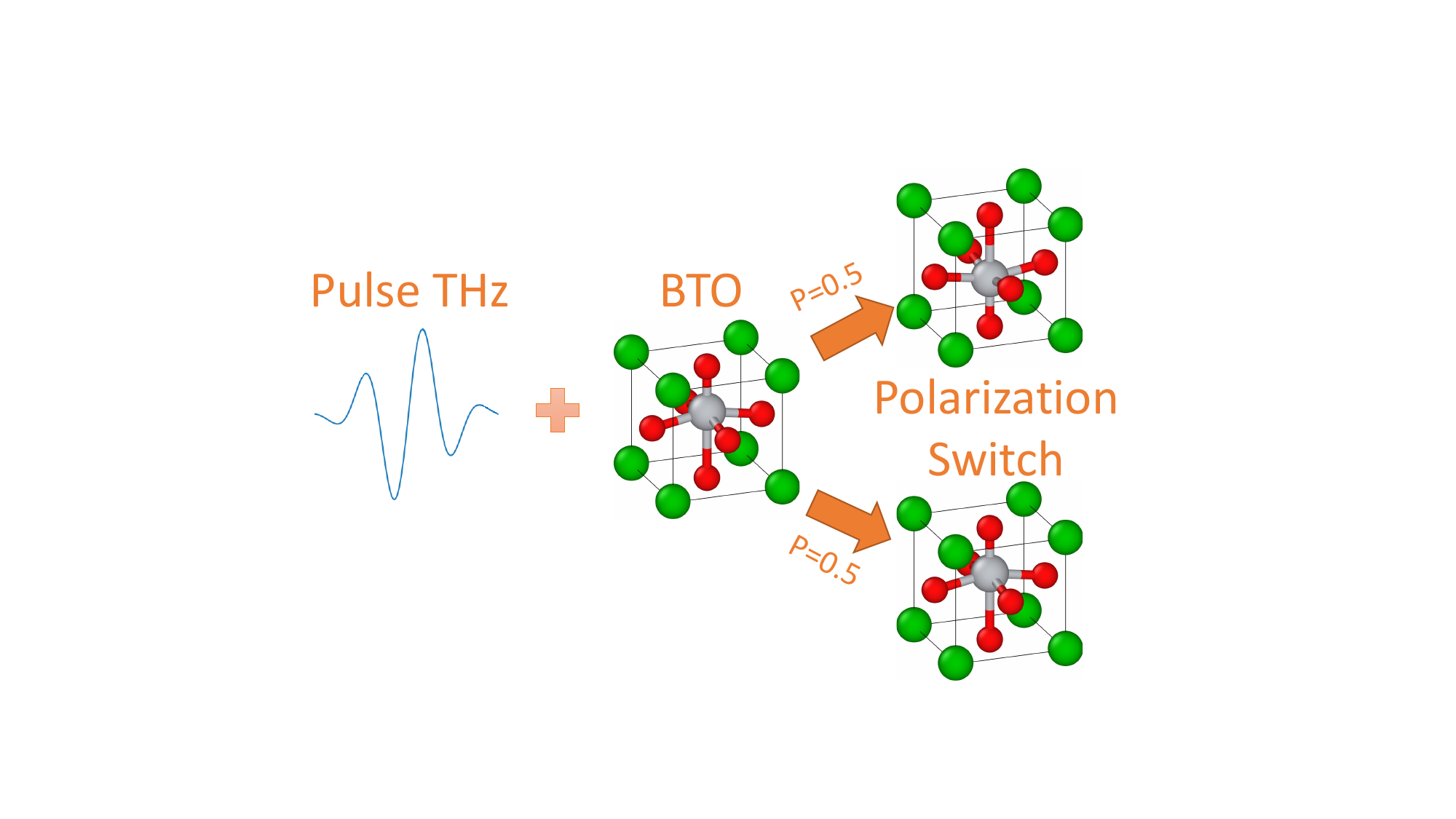}
\end{center}
\end{tocentry}

\begin{abstract}
The challenge of achieving ultrafast switching of electric polarization in ferroelectric materials remains unsolved, as there is no experimental evidence of such switching to date. In this study, we have developed an enhanced model that describes switching within a two-dimensional space of generalized coordinates at THz pulses. Our findings indicate that stable switching in barium titanate cannot be achieved through a single linearly polarized pulse. When the intensity of the linearly polarized pulse reaches a certain threshold, the sample experiences depolarization, but not stable switching. Our study also reveals that phonon friction plays a minor role in the switching dynamics and provides an estimate of the optimal parameters of the perturbing pulse with the lowest intensity that results in depolarization of an initially polarized sample.
\end{abstract}

\section{Introduction}
Developing non-volatile memory devices with fast writing and reading operations while minimizing power consumption is a challenge in information storage. However, traditional magnetic storage and flash may not be suitable for future fast devices due to their limited operation speed, which is in the milliseconds range. Thus, this challenge can only be addressed by utilizing different physical mechanisms for writing and reading bits.

A potential physical mechanism for write operation is magnetization switching by an ultra-short electromagnetic pulse of optical or THz range. This mechanism has shown promise in previous studies~\cite{kirilyuk2010ultrafast,abdullaev2020ferroelectric, kimel2020fundamentals}.
Similarly, electric fields can be utilized for ultra-fast polarization switching in ferroelectric materials.
Although this possibility has garnered significant attention, it has not yet been observed experimentally. 
The closest successful result to date, which involved reversible polarization change, was achieved by Mankowsky \textit{et al.} in their work on lithium niobate~\cite{mankowsky2017ultrafast}.
Other studies~\cite{brekhov2015photoinduced,chen2016ultrafast, brekhov2018optical,brekhov2019temperature,grishunin2019transient,bilyk2021transient} have also explored the selective excitation of lattice vibrations under ultra-short optical or THz pulses, which is essential for achieving practical polarization switching.

The absence of a predictive model poses a significant obstacle to experimentally observing ultra-fast switching of electric polarization.
Such a model could provide optimal pulse parameters and answer a series of questions, such as: which normal mode should receive energy injection, whether energy should be injected directly into the mode that leads to switching or another strongly coupled mode; whether it is beneficial to use a series of pulses; which pulse polarization is optimal for switching; whether pulse shape affects switching; and which ferroelectric material is best suited for ultra-fast switching of electric polarization, among others.

In this research, we improved and tested a theoretical model for ultra-fast polarization switching, which has previously been proposed in various studies~\cite{qi2009collective, subedi2015proposal, mankowsky2017ultrafast, zhilyaev2023modelling}.
To calculate material constants of ferroelectrics as  oxides and chalcogenides, first principles methods like Density Functional Theory (DFT) are often utilized~\cite{Waghmare_2014}. 
These methods are effective in determining the structure of stable polarized states, energy barriers, ions' effective charges, polarization values, and the phonon spectrum~\cite{shin2005development,rabe2007first,stroppa2015ferroelectric,subedi2015proposal,smidt2020automatically, Tantardin2023,Semenok2022}.
Moreover, it is important to highlight that DFT calculations' results are highly dependent on the chosen exchange-correlation functional~\cite{zhang2017comparative}.
Classical molecular dynamics (MD) simulations enable the examination of ultra-fast polarization switching at an atomistic level~\cite{qi2009collective} and even take into account domain behavior~\cite{boddu2017molecular}.

The proposed model  aims to investigate ultra-fast polarization switching in ferroelectrics.
The model utilizes a system of ordinary differential equations (ODEs) to represent the time progression of the generalized coordinates within a ferroelectric material's elementary cell.
Radiation interaction is included by incorporating a perturbation force within the ODE, which functions for a specific duration.
The potential energy surface (PES) is obtained from DFT calculations.
Barium titanate (BTO) is used as a test material in this research, as it is a well-studied, prototypical ferroelectric material.

The proposed model primarily builds upon earlier works~\cite{subedi2015proposal,fechner2016effects,juraschek2017ultrafast, subedi2017midinfrared,itin2017efficient}, where a similar approach was employed for polarization switching and structure changes driven by ultra-short pulses. However, two significant modifications were introduced.
First, instead of representing the PES in the form of Taylor's series, we directly interpolate PES using cubic splines.
This is because switching results in substantial atomic displacement, leading to high numerical errors in Taylor's series.
Second, in terms of generalized coordinates, we consider the polarization mode ($q_p$), which undergoes the switch, and the normal mode ($Q_{IR}$) where radiation is pumped. In contrast to previous studies~\cite{subedi2015proposal}, both generalized coordinates were normal modes, this approach contradicts the fact that the potential must be scalar, independent of the crystal's symmetry (for more details, please refer to~\cite{mertelj2019comment}).
The article is structured as follows. In the methods section, we give details of calculating the PES and constructing the system of ODEs.
The results and discussion section presents the data obtained for BTO, along with a discussion on metastable switching, effective friction, perturbation duration, and optimal frequency.
The conclusion section provides general observations and recommendations for future experiments.

\section{Computational Details}
We take the experimental values of a material's unit cell and relaxing the atomic positions to obtain the equilibrium structure. Both ionic relax and calculations for phonon spectra and energies are carried out using the Vienna Ab initio Simulation Package (VASP) software package~\cite{kresse1994ab,kresse1994norm,kresse1996efficiency,kresse1996efficient}, employing a plane-wave basis set.
The projector augmented-wave (PAW) pseudopotential with a general gradient approximation PBE~\cite{perdew1996generalized} and a cutoff energy of 600 \textit{eV} is utilized in all calculations.
Numerical integration over the Brillouin zone is conducted using an \hbox{$8\times 8\times8$} k-point sampling with a Gamma-centered grid.

The phonon dispersion curves are calculated within the framework of Finite Displacements (FD) using the Phonopy code~\cite{phonopy}.
All corresponding DFT calculations are executed for a perfect \hbox{$2\times2\times2$} supercell structure.
After identifying the normal modes, the PES is calculated as a function of two independent normal mode generalized coordinates: $q_{p}$ (polarization mode) and $Q_{IR}$ (high-frequency mode).

The individual atomic displacements, associated with the generalized coordinate $q_\text{p}$, can be expressed as:
\begin{equation}
    U_i = \left(\frac{q_{p} + 1}{2}\right)(Z_i^{D} - Z_i^{U}) + Z_i^U
    \label{eq:dis_polarization}
\end{equation}
Here, $U_i$ represents the displacement of the $i$-th atom, while $Z_{i}^{U}$ and $Z_{i}^{D}$ denote the coordinates of the $i$-th atom in the direction of polarization, corresponding to equilibrium positions with positive and negative polarization, respectively.

The individual atomic displacements, related to the generalized coordinate $Q_{IR}$ are given by:
\begin{equation}
    U_i = \frac{Q_{IR}}{\sqrt{m_i}}\eta^{IR}_i
    \label{eq:disp_norm_mode}
\end{equation}
where $U_i$ is the displacement of a \textit{i}-th atom, $m_{i}$ – atomic mass,  $\eta_i^{IR}$ is the corresponding component of the normal mode dimensionless eigenvector. 

\begin{figure}
\centering
\includegraphics[width=0.50\textwidth]{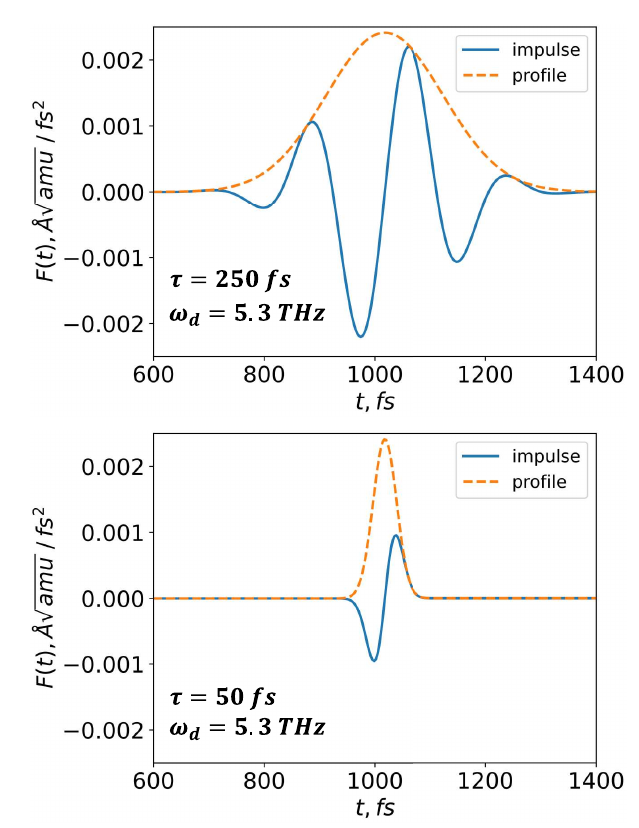}
\caption{A visual representation of the perturbing force $F(t)$ is shown for two pulse durations (250 and 50 \textit{fs}) and a frequency related to the high-frequency optical normal mode (5.3~THz).
It is essential to note that the frequency is significantly high, allowing approximately 2 oscillations to fit within the 250 \textit{fs} envelope.}\label{fig:impulse}
\end{figure}   

The PES is interpolated using cubic splines~\cite{2020SciPy-NMeth} at points where DFT calculations are obtained, allowing us to define the PES continuously as $V(q_{p}, Q_{IR})$.
The dynamic behavior of the coupled generalized coordinates is characterized by a system of associated nonlinear differential equations of motion:
\begin{equation}\label{eq:cos}
\begin{aligned}
\ddot q_{p} + \gamma \dot q_{p} = & -\frac{\partial V(q_{p}, Q_{IR})}{\partial q_{p}} \\
\ddot Q_\text{IR} + \gamma \dot Q_{IR} = & -\frac{\partial V(q_{p}, Q_{IR})}{\partial Q_{IR}} + F(t)
\end{aligned}
\end{equation}

where $\gamma$ represents the effective friction coefficient, and $F(t)$ is the initial force exerted on the system due to external pulse perturbation.
The integration of Eq.~\eqref{eq:cos} is performed using the \verb|odeint| library from the SciPy package~\cite{2020SciPy-NMeth}.
We assume $F(t)$ takes the following form:
\begin{equation}\label{eq:force}
F(t) = F_{0}\:\sin(\omega_{d} t)\: \exp\left[-4ln2 \: \left(\frac{t^2}{\tau^2} \right) \right] 
\end{equation}
where $F_{0}$ is the force amplitude, $\omega_{d}$ is the perturbation's driving frequency (assumed to equal $\omega_{IR}$, unless stated otherwise), and $\tau$ is the pulse's time length. A graphical representation how the the perturbing force increases with increasing of pulse duration is illustrated in Fig.~\ref{fig:impulse}.

\begin{figure}
\centering
\includegraphics[width=0.8\textwidth]{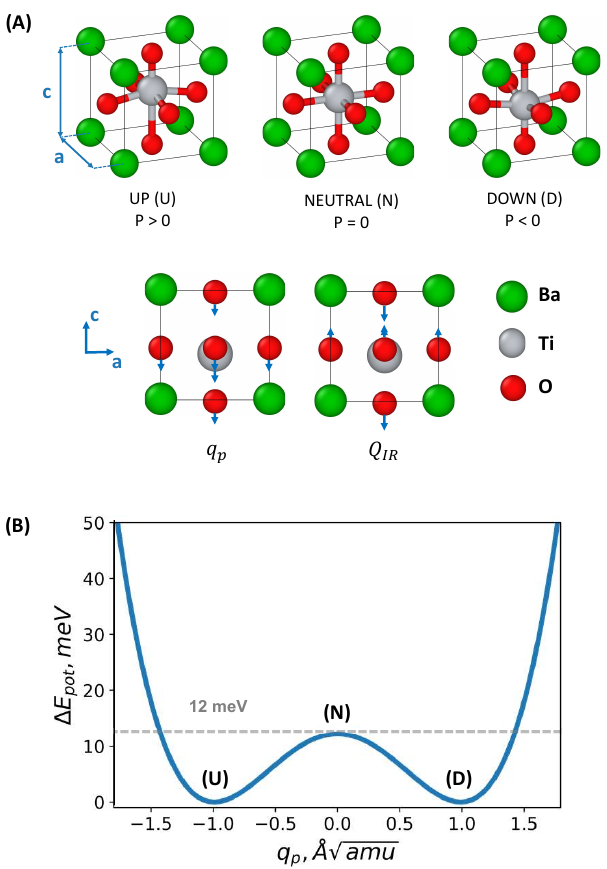}
\caption{(\textbf{A}) An atomic illustration of the tetragonal ferroelectric phase in barium titanate (BTO) \textit{P4mm} is provided. Primarily, electric polarization switching is linked to the motion of the titanium atom along the c-axis: UP (U), same direction to the c-axis; NEUTRAL (N), no polarization; DOWN (D), opposite direction to the c-axis. The figure also illustrates the displacement patterns of the generalized coordinates denoted by $q_p$ and $Q_{IR}$.(\textbf{B}) The energy barrier for BTO divides the two stable states related to the nominal downward and upward electrical polarization. The barrier's height, as calculated from first principles calculations, and it is approximately 12 \textit{meV}, which agrees well with results from similar studies.}
\label{fig:barrier_structure}
\end{figure} 

\section{Results and Discussion}
The ferroelectric state of BTO is present in the crystal structure featuring a lattice with the \textit{P4mm} space group. We adopt the following experimental crystal unit cell parameters: $a = 3.986$~\AA{} and $c = 4.026$~\AA~\cite{shirane1957neutron}.
The primitive unit cell is composed of one barium atom, one titanium atom, and three oxygen atoms (refer to Fig.~\ref{fig:barrier_structure}).
This structure gives rise to 15 normal modes at the Gamma point, including three acoustical and twelve optical branches, which are of particular interest to us.
The optical normal modes at the gamma point can be decomposed as $\Gamma = 3A_1 + B_1 + 4E$.
The initial cubic symmetry \textit{Pm-3m} of the paraelectric BTO crystal at 130 Celsius goes for transition to ferroelectric state through atomic displacements strictly along the c-axis into tetragonal \textit{P4mm} symmetry \cite{Smith_2008}.
Consequently, the coupling between normal modes and the motion ($q_p$) responsible for polarization switching is likely to occur with normal modes that possess large c-axis components in their eigenvectors.
In BTO, these modes are 5, 9, and 11, corresponding to frequencies of 5.3, 8.8, and 14.1 THz.
The excitation of only three low frequency modes allowed us to avoid the nonlinear coupling between low and high frequency modes that it is known to affect the polarization switching in such material when both are present \cite{Subedi_2015}.
In this work, we chose to investigate mode 5 because it represents a typical frequency that can be achieved with modern powerful terahertz radiation sources avoiding the presence of second harmonics \cite{mankowsky2017ultrafast}.

\begin{figure}
\centering
\includegraphics[width=0.50\textwidth]{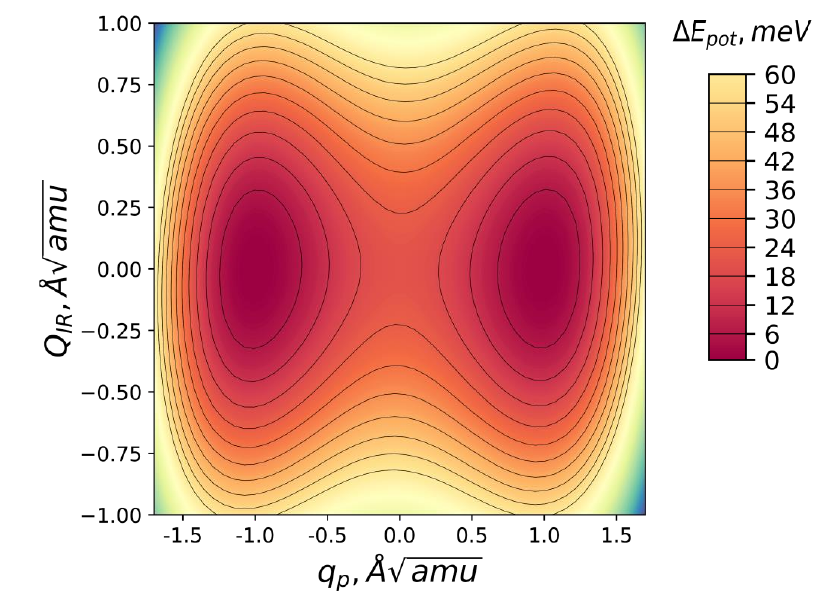}
\caption{Barium titanate potential energy surface is illustrated in generalized coordinates ($q_p, Q_{IR}$).
The heat map displays energy in \textit{eV} units, measured from the base value of potential energy at (0, 0).}
\label{fig:pes}
\end{figure}   
\unskip

The PES was computed in the space of two generalized coordinates ($q_p$, $Q_{IR}$), with each representing collective displacement of all atoms in the unit cell (refer to eq.~\ref{eq:dis_polarization} and~\ref{eq:disp_norm_mode}).
The sampling for $q_p$ was performed in the range from -2.0 to 2.0 with a step of 0.05 in \r{A} $\sqrt{amu}$, while the sampling for $Q_{IR}$ was carried out in the range from -3.0 to 3.0 with a step of 0.01 in \r{A} $\sqrt{amu}$ units, resulting in a total of 48000 static DFT calculations. 
The point representation of PES was interpolated using cubic splines for solving the systems of ODEs.
This method offers a more accurate representation of polarization switching compared to the Taylor series expansion, which is only effective in the local vicinity of the expansion point\cite{subedi2015proposal, mankowsky2017ultrafast}.

A PES cross-section (shown in Fig.~\ref{fig:pes}) along the direction $Q_{IR} \sim 0 \:$ \r{A} $\sqrt{amu}$ enables the examination of the barrier obtained by linearly interpolating the system's atomic coordinates from an upward polarization state to a downward polarization state.
For DFT calculations, the barrier height is found to be $\sim 12$ \textit{meV}, which is consistent with other calculations employing the PBE exchange-correlation potential ~\cite{zhang2017comparative}.

\begin{figure}
\centering
\includegraphics[width=1\textwidth]{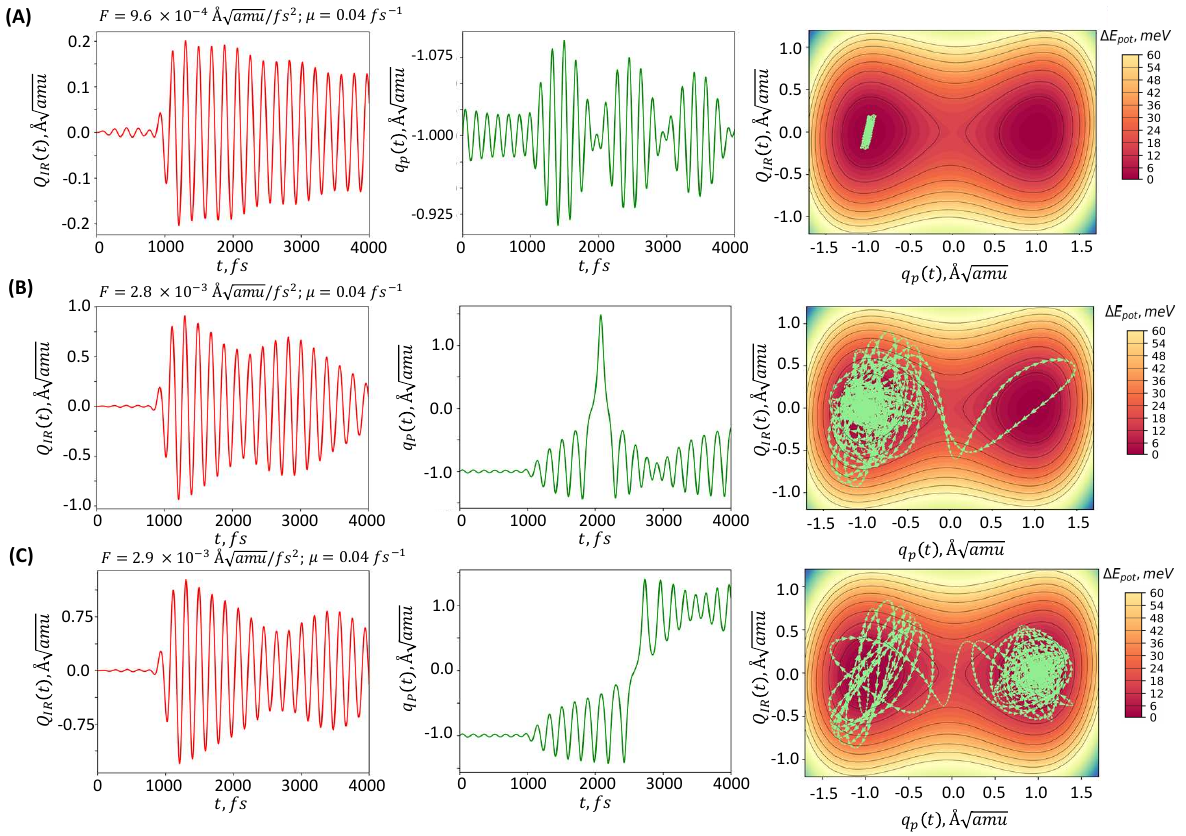}
\caption{The time evolution of generalized coordinates under the influence of varying pulse amplitudes is depicted, with the trajectory of generalized coordinates on the potential energy surface shown as a green line. (A) When the perturbation amplitude is relatively small, no switching takes place, and the system remains at its initial minimum; (B) The switching may not be "stable" - the system can momentarily enter a state with opposite electrical polarization, but due to inertia, it may return to and remain in the initial minimum, preventing the switching from taking place; (C) If the perturbation amplitude is large enough, switching occurs, and the system transitions into a state with reversed electric polarization.}
\label{fig:switching_scenario}
\end{figure}

To analyze the trajectory of generalized coordinates under a perturbing pulse for differing perturbation amplitudes, a series of calculations was performed (refer to Fig.~\ref{fig:switching_scenario}).
The effective friction coefficient was set at $\mu = 0.04$ $fs^{-1}$. Three distinct scenarios were observed: 

\begin{enumerate}
    \item When the perturbation force is not sufficient, the system remains in the initial minimum, with the trajectory localized nearby (see Fig.~\ref{fig:switching_scenario}A).
    \item A scenario not typically addressed by other authors~\cite{subedi2015proposal, mankowsky2017ultrafast}, but worth noting, involves the system entering a different polarization state only to return to its initial state after a period of time due to inertia. Thus, even a strong enough perturbation impulse may not alter the final electric polarization (see Fig.~\ref{fig:switching_scenario}B).
    \item Upon reaching a specific threshold for perturbation amplitude, enough energy is transferred into the system to surpass the barrier between local minima, causing the system to switch to a state with reversed polarization (see Fig.~\ref{fig:switching_scenario}C).
\end{enumerate}

A reversible polarization switch was previously observed in a study~\cite{qi2009collective} where lead titanate (PTO) was modeled at the atomic level.
Therefore, exposing BTO to a single polarization pulse could lead to irreversible switching if the pulse parameters fall within a narrow range. 
However, even with carefully chosen pulse parameters, irreversible polarization switching might not be achieved due to the chaotic nature of polarization switching~\cite{chirikov1979universal}.
Further research is needed to investigate this hypothesis in detail.

\begin{figure}
\centering
\includegraphics[width=1\textwidth]{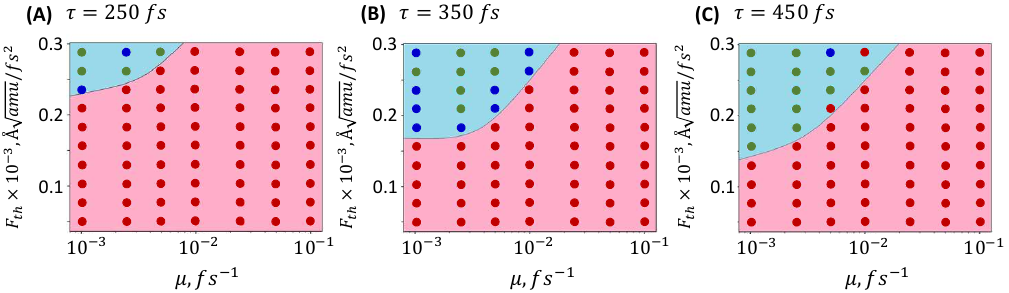}
\caption{A series of computations were performed for threshold amplitude of the perturbing pulse $(F_{th})$ from 0.0 to 0.3 \r{A} $\sqrt{amu}$ / $fs^{2}$ at three different pulse lengths: (A) 250 \textit{fs}, (B) 350 \textit{fs}, (C) 400 \textit{fs}. For each computation set, the friction coefficient $(\mu)$ was modified over a wide range of values, from $10^{-3}$ to $10^{-1}$ $fs^{-1}$. In each calculation, the presence or absence of polarization change was noted: red circles represent calculations where polarization switching did not occur; blue circles indicate instances where polarization shifted but eventually returned to its original state; and green circles denote calculations where the polarization switched to its opposite value.}
\label{fig:friction}
\end{figure}  

A crucial fitting parameter in the equations that describe the dynamics of generalized coordinates is the friction coefficient.
Estimating this coefficient can be done through calibration experiments. 
Nonetheless, several factors can impact the friction coefficient, such as: (1) the domain structure's dependency on geometrical dimensions of the ferroelectric material sample; (2) the influence of neighboring unit cells (not considered in this work); (3) the density of local defects.
As a result, we conducted calculations by varying the friction coefficient over a broad range, analyzing its influence on the threshold switching force and switching stability (refer to Fig.~\ref{fig:friction}).
Calculations were performed for three pulse duration: 250, 350, and 450 \textit{fs}, and a set of friction coefficients ranging from $10^{-3}$ to $10^{-1}$ $fs^{-1}$.
The calculations determined whether a switch occurred and if it was reversible or irreversible.

We observed that different pulse length only with friction coefficient between $10^{-3}$ and $10^{-2}$ $fs$ generates switching of polarization. The increasing of pulse length allowed us to observe a decreasing of necessary amplitude of the perturbing pulse $(F_{th})$ to observe switching of polarization.

\begin{figure}
\centering
\includegraphics[width=0.50\textwidth]{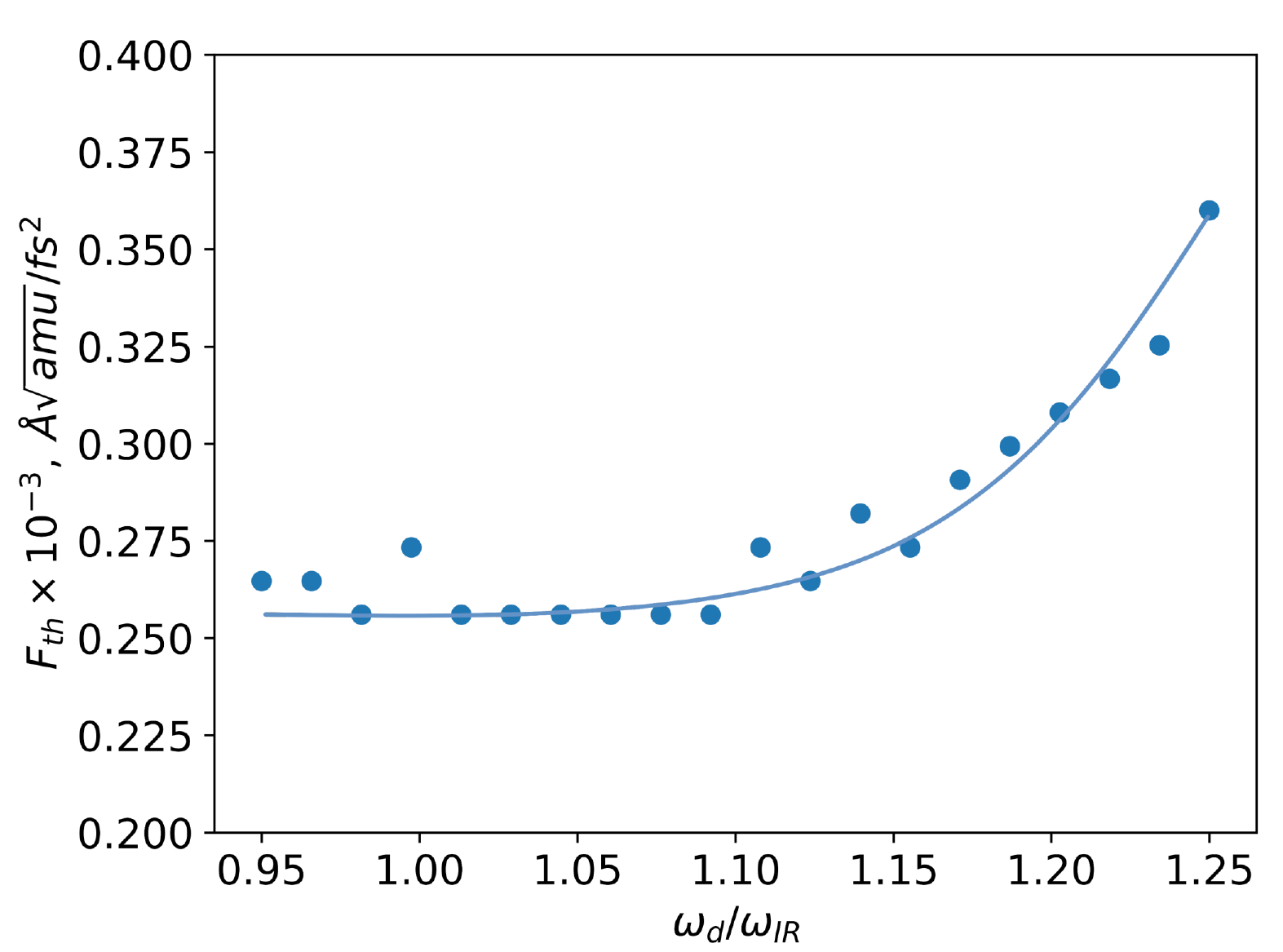}
\caption{The relationship between the amplitude of the perturbing pulse, which causes switching, and the frequency of the perturbing pulse is demonstrated. 
The graph indicates that the lowest threshold force amplitude falls within the range of $0.95~\omega_{IR}$ to $1.10~\omega_{IR}$. A continuous line is included merely to serve as a visual guide. Pulse duration 250 $fs$.}
\label{fig:omega}
\end{figure}   

Additionally, the frequency of the perturbing pulse was varied in the calculations (see Fig.~\ref{fig:omega}).
The lowest threshold amplitude was observed for frequencies in the range of 0.95 $\omega_{IR}$ to 1.10 $\omega_{IR}$ for the eigenfrequency of the perturbed $Q_{IR}$ mode, while the threshold force amplitude reduction was approximately 1.6 times.
The optimal frequency shift observed results from coupling with the high-frequency mode and the presence of a friction term in the equation of motion. 

An analysis of the motion equation reveals that for small amplitude excitations~\cite{mankowsky2015coherent, fechner2016effects}, coupling effects cause renormalization of the optimal frequency, $\omega_{IR}$.
A frequency shift in an underdamped oscillator is a well-studied phenomenon~\cite{hayek2003mechanical}.

Let's also estimate the fluence corresponding to a typical force at which polarization switching occurs. We adopt the smallest noted value (see Fig.~\ref{fig:omega}), which is on the order of $F_{th} = 2.5 \times 10^{-4} ~\AA \sqrt{amu} / fs^2$.
This force $(F_{th})$ is equivalent to the acceleration $a_{th} = F_{th} / \sqrt{m_{Ba^{+}}} = 1.7 \times 10^{-5}~\AA / fs^2$, which represents the acceleration of the $Ba^{+}$ ion created by the electric field $E_{th} = m_{Ba^{+}} \cdot a_{th} / q_{Ba^{+}} = 1.2 \times 10^{11}~V$.
Subsequently, the energy density of such a field is linked to fluence $W = \epsilon_{0}E^2 / 2 \cdot \Omega = F_{th} \cdot S$, which infers $F_{th} = \epsilon_{0} \cdot E^2/2 \cdot h$, where $\Omega$, $S$, and $h$ represent the volume, surface area, and the length of the unit cell in the 'c' direction (see Fig~\ref{fig:barrier_structure}a) respectively, and $\epsilon_0$ is the vacuum permittivity.
This simple estimation yields a value of $F_{th} = 250~mJ/cm^2$.
Although this is a rather basic analysis, the derived estimation should be approached with caution.
For comparison in article~\cite{mankowsky2017ultrafast} studying lithium niobate (LNO), the onset of polarization switching occurred at fluences of 95 $mJ/cm^{2}$, which is on the same order of magnitude as our $F_{th}$ estimate for BTO.

\section{Conclusions}
In this study, a model was examined and evaluated to characterize the ultrafast switching polarization in ferroelectric materials using BTO as test case.
Analyzing the proposed model indicates that exist an operative range of the friction coefficient where the ultrafast switching polarization has the highest probability to happen. 
Such probability increases with increasing of pulse and the smallest threshold force amplitude necessary for switching is achieved within the range of $0.95~\omega_{IR}$ to $1.10~\omega_{IR}$, where $\omega_{IR}$ represents the normal mode frequency.
Polarization switching has been shown to be reversible, and it is probably a random process, meaning that slight changes in the perturbing pulse parameters might lead to an opposite final polarization.
Thus, the complexity of the model in the future should include arbitrary polarization of the perturbing pulse, which may prove difficult to interpret, and possibility to consider multi-pulse cases. 
For example involving the depolarization potential, which is generated by secondary high-frequency pulses, which inject energy into the electronic subsystem raising the electronic temperature to tens of \textit{eV} favoring the switching of polarization as seen in previously works\cite{Abalmasov_2020,mankowsky2017ultrafast,ernstorfer2009formation}.

\section{Data Availability}
All data generated or analyzed during this study are included in this published article and supplementary data are available by request addressed to the corresponding authors.

\begin{acknowledgement}
This work was supported by the Russian Science Foundation grant number 20-72-10178 and Russian Academy of Sciences project number 121032500059-4.
The computations were carried out on supercomputer MVS-10Q at Joint Supercomputer Center of the Russian Academy of Sciences (JSCC RAS), the supercomputer Zhores (CDISE, Skoltech, Russia)~\cite{zacharov2019zhores}, and Skoltech HPC cluster “ARKUDA”.
\end{acknowledgement}


\bibliography{acs-achemso}

\end{document}